\documentclass[preprintnumbers, floatfix, preprintnumbers, letterpaper, superscriptaddress,nofootinbib]{revtex4}
\usepackage{graphicx}
\usepackage{microtype}
\usepackage{amsmath}
\usepackage{amssymb}
\usepackage{subfigure}
\usepackage{hyperref}
\usepackage{url}
\usepackage{xcolor}
\usepackage{color}
\usepackage{mathrsfs}
\usepackage{calrsfs}
\usepackage{amsfonts}
\usepackage{eufrak}
\usepackage{tabularx}
\usepackage{eucal}
\usepackage{latexsym}
\usepackage{ragged2e}
\usepackage{epsfig}
\usepackage{textcomp}

\usepackage{caption}
\DeclareCaptionJustification{justified}{\leftskip=0pt \rightskip=0pt \parfillskip=0pt plus 1fil}
\definecolor{vividviolet}{rgb}{0.62, 0.0, 1.0}
\definecolor{amaranth}{rgb}{0.9, 0.17, 0.31}
\definecolor{palatinateblue}{rgb}{0.15, 0.23, 0.89}
\definecolor{brightpink}{rgb}{1.0, 0.0, 0.5}
\definecolor{cornflowerblue}{rgb}{0.39, 0.58, 0.93}
\definecolor{deepcarminepink}{rgb}{0.94, 0.19, 0.22}
\definecolor{radicalred}{rgb}{1.0, 0.21, 0.37}

\hypersetup{ linktoc=all,
    colorlinks, linkcolor={palatinateblue},
    citecolor={brightpink}, urlcolor={amaranth}
}

\graphicspath{{Images/}}

\renewcommand{\d}[1]{\ensuremath{\operatorname{d}\!{#1}}}

\graphicspath{{Images/}}
\renewcommand{\d}[1]{\ensuremath{\operatorname{d}\!{#1}}}
\makeatletter
\def\@fnsymbol#1{\ensuremath{\ifcase#1\or \ddagger \or  $\textleaf$  \or \dagger
\else\@ctrerr\fi}}%

\makeatother

\def\sideremark#1{\ifvmode\leavevmode\fi\vadjust{\vbox to0pt{\vss
 \hbox to 0pt{\hskip\hsize\hskip1em
 \vbox{\hsize1.3cm\tiny\raggedright\pretolerance10000
 \noindent #1\hfill}\hss}\vbox to8pt{\vfil}\vss}}}%
\def\beq{\begin{equation}}
\def\eeq{\end{equation}}

\begin{document}

\title{Thermodynamics of Shearing Massless Scalar Field Spacetimes is Inconsistent \\With the Weyl Curvature Hypothesis}

\author{Daniele \surname{Gregoris}}
\email{danielegregoris@libero.it}
\affiliation{Center for Gravitation and Cosmology, College of Physical Science and Technology, Yangzhou University, \\180 Siwangting Road, Yangzhou City, Jiangsu Province, P.R. China 225002}
\affiliation{School of Aeronautics and Astronautics, Shanghai Jiao Tong University, Shanghai 200240, China}

\author{Yen Chin \surname{Ong}}
\email{ycong@yzu.edu.cn}
\affiliation{Center for Gravitation and Cosmology, College of Physical Science and Technology, Yangzhou University, \\180 Siwangting Road, Yangzhou City, Jiangsu Province, P.R. China 225002}
\affiliation{School of Aeronautics and Astronautics, Shanghai Jiao Tong University, Shanghai 200240, China}

\author{Bin \surname{Wang}}
\email{wang\_b@sjtu.edu.cn}
\affiliation{Center for Gravitation and Cosmology, College of Physical Science and Technology, Yangzhou University, \\180 Siwangting Road, Yangzhou City, Jiangsu Province, P.R. China 225002}
\affiliation{School of Aeronautics and Astronautics, Shanghai Jiao Tong University, Shanghai 200240, China}

\begin{abstract}
Our Universe has an arrow of time. In accordance with the second law of thermodynamics, entropy has been increasing ever since the Big Bang. The fact that matter is in thermal equilibrium in the very early Universe, as indicated by the cosmic microwave background, has led to the idea that gravitational entropy must be very low in the beginning. Penrose proposed that gravitational entropy can be quantified by the Weyl curvature, which increases as structures formed. A concrete realization of such a measure is the Clifton-Ellis-Tavakol gravitational entropy, which has been shown to be increasing in quite a number of cosmological models. In this work, we show a counter-example involving a class of inhomogeneous universes that are supported by a chameleon massless scalar field and exhibit anisotropic spacetime shearing effects.  In fact, in our model the Clifton-Ellis-Tavakol gravitational entropy is increasing although the magnitude of the Weyl curvature is decreasing; this is due to the growth of the spacetime shear. The topology and the values of the three free parameters of the model are constrained by imposing a positive energy density for the cosmic fluid, and the thermodynamical requirements which follow from the cosmological holographic principle and the second law. It is shown that a negative deceleration parameter and a time decreasing Weyl curvature automatically follow from those conditions.  Thus, we argue that our model can account for the formation of some primordial structures, like the Large Quasar Groups, which has required a non-standard evolution of the spatial anisotropies.

\end{abstract}
\maketitle

\section{Introduction: The Arrow of Time and Gravitational Entropy}

The notion of time has always been an intriguing subject in both science and philosophy. Time, unlike space, has a \textit{direction}, it inevitably ``flows'' from the past to the future. In terms of the second law of thermodynamics, the arrow of time is reflected in the fact that statistically speaking, entropy tends to increase rather than decrease. This is because, as Boltzmann and Gibbs tell us, entropy counts the number of microstates via the formula $S=k_B\log W$, where $W$ is the number of microstates in which the energy of the molecules in a system can be arranged. In the phase space then, a system naturally evolves from a region of smaller phase space volume to one with larger volume. There is hardly any physics at this stage -- what we are dealing with is combinatorics. For example, there are just a lot more ways to have the wires of an earphone all tangled up than not. Therefore it is not at all surprising that one expects entropy to increase merely because there are more ways for the configurations of a closed system to be in high entropy states than in lower entropy ones. The surprising thing is that this argument is time symmetric and so by appealing to combinatorics alone we should also expect entropy to be increasing towards the past. The fact that it does not -- for otherwise there would cease to be an arrow of time -- 
means that the beginning of the Universe must have a very low entropy in some sense.
In other words, the second law tells us that, since entropy is increasing, it must have been lower in the past, all the way back to the Big Bang (see, however, \cite{1602.05601}). It is because of the very low entropy of the very early Universe that we exist at all, if everything has been in equilibrium at the very beginning, nothing would have happened.

The physics thus comes in by demanding that the \textit{initial condition} of the Universe must be such that it is at a very low entropy state, and then the combinatorics nature of the second law takes over and naturally evolves it towards a higher entropy future, governed by various laws of physics. The question is not why entropy increases, as that was settled by Boltzmann already. The question is: \textit{why} is the very early Universe at such a low entropy state? In other words, the problem of the arrow of time \textit{is} the problem of the initial condition of the Universe. To quote Richard Feynman in his \emph{Lectures on Physics} \cite{feynman},
``so far as we know, all the fundamental laws of physics, such as Newton's equations, are reversible. Then where does irreversibility come from? It comes from order going to disorder, but we do not understand this until we know the origin of the order.''\footnote{For more detailed discussions regarding the issue of the arrow of time and its cosmological origin, see \cite{Price1, Price2, vaas}. For an introductory article, see \cite{Price3}.} We know from observational data of the cosmic microwave background (CMB) that the matter at the end of the epoch of recombination was already at thermal equilibrium, as shown by the almost perfect Planck distribution of the CMB. Normally we associate thermal equilibrium as a high entropy state. Thus, in order to have a low \emph{total} entropy back then, the gravitational entropy must be properly taken into account. Indeed, a smoothly distributed matter field like the conditions in the early Universe (with the density perturbation being as mere $\delta \rho/\rho \sim 10^{-5}$) is a low entropy state as far as gravity is concerned -- gravity tends to clump and contract matter, so structure formation is in accordance with the second law. 

How then does one define or quantify gravitational entropy? Penrose proposed that Weyl curvature can be used for this exact purpose \cite{wcc1}. Indeed, Weyl curvature describes how the shape of a body is distorted by the ``tidal force'' of a gravitational field \cite{0103044}. It tends to increase during structure formation and gravitational collapse. Penrose thus proposed the \emph{Weyl curvature hypothesis}, which claims that near the past singularity (the Big Bang) the Weyl curvature must vanish, and then it started to rise monotonically thereafter as matter starts clumping, stars and galaxies forming, and so on. If there is a crunch, future singularity can be arbitrarily distorted and so has large Weyl curvature, unlike the initial singularity. A concrete realization of the notion of gravitational entropy is the Clifton-Ellis-Tavakol (CET) entropy \cite{form7}, which essentially also measures the Weyl curvature.

The question of the arrow of time is a deep one: there have been many proposals in the literature that attempt to explain why the initial gravitational entropy is so low, 
including -- but not limited to -- weakening the strength of gravity during the very early universe (so that the smooth initial state is not an unusually low entropy state) \cite{0911.0693}, constructing a time symmetric universe\footnote{This is so that it manifestly passes the ``double standard test'' made explicit by Price \cite{Price1}: if a physical mechanism is supposed to explain past low entropy (which gives rise to what we experience as the passing of time) without itself sneaking in time asymmetry, then that mechanism should also be applicable to the  ``end state of time'', i.e. future conditions. In other words the scenario must make sense if we reverse the arrow of time, \textit{unless} there is some \textit{a priori} ``natural'' reason that breaks the symmetry and makes the past objectively distinct from the future (if so, one should explain this).} (this ranges from the early model of Gold \cite{gold}, in which the entropy gets lower in the future as the universe shrinks in size, to a more sophisticated model of time symmetric \emph{multiverse} by Carroll and Chen \cite{0410270v1}; see also \cite{0111191, 0301042,0712.0571, 1305.3836}), Penrose's conformal cyclic universe \cite{cyclic}, and ``creation on a torus'' in stringy model that identifies gravitational entropy with some notion of ``geometric entropy'' \cite{0611088v3, 0711.1656v2}. There is as yet no consensus to the solution of the arrow of time problem, and it is not our aim in this paper to provide a better explanation.

Instead, we are interested in a more modest question: \emph{Is the Weyl curvature hypothesis correct? More specifically, does Weyl curvature always increase in any physically realistic universe?} By physically realistic we do not mean that it must satisfy all the observational data of the actual Universe\footnote{We have reserved capital letter ``Universe'' for the actual one we live in, while lower case ``universe'' refers to any generic universe. Of course, some statements regarding the latter might turn out to hold also for the former.}, but only the weaker requirement that it should satisfy well-established laws of physics in general, and notably the laws of thermodynamics in particular. If the Weyl curvature hypothesis is indeed correct, then it should hold in any logically consistent universe with a thermodynamical arrow of time, as structures are formed. Structure formation comes in the form of inhomogeneities and anisotropies. In this work we investigate the joint effect of spatial inhomogeneities and of a  cosmological shear, and we constrain the model at the theoretical level by imposing a few physical conditions: the cosmic fluid in the model must have a positive energy density, the second law of thermodynamics must be obeyed in the matter sector, and the total matter entropy must be bounded by the area of the dynamical apparent horizon (the ``cosmological holographic principle'', see more below) \cite{bousso}. We then show that in this specific example -- \emph{despite all these physically realistic requirements} -- the Weyl curvature hypothesis does \emph{not} hold -- the Weyl curvature is monotonically decreasing (bf however this would not be the case for the CET gravitational entropy), while spacetime shear continues to increase as the universe expands. 

Let us now move on to explain the shearing spacetime cosmological model, before returning to the issue of gravitational entropy. Our paper is organized as follows. In Sec. (\ref{sect1}) we will provide further motivation to the model from the viewpoint of cosmology, irrespective of the arrow of time issue. In other words, the model we examined is not an exotic one cooked up just to serve as an ad hoc counter-example to the Weyl curvature hypothesis, but has physical motivation on its own. In Sec. (\ref{sect2}) we shall review the most important physical properties of the model under analysis by computing its cosmological parameters. Then in Sec. (\ref{sect3})  we shall exhibit the constraints on the parameters of the model, which are derived from the cosmological holographic principle and from the second law of thermodynamics. We also clarify the role of the position of the observer in such a universe. In Sec. (\ref{sect4}) we shall return to compute the gravitational entropy, as defined by the CET proposal,  and we will explain why our cosmological model may account for the existence of some exotic astrophysical structures, like the Large Quasar Groups, whose sizes are larger than the homogeneity scale assumed by the standard model of cosmology.
In the concluding Sec. (\ref{sect5}), we will discuss the notion of gravitational entropy in general and what our finding might imply in the larger context of the arrow of time problem. In addition, we shall put our work in the context of the current research developments in theoretical cosmology, which are gradually appreciating the importance of constructing model-independent (i.e. not relying on the Copernican principle in any step of the analysis) techniques for constraining cosmological models.

\section{Shearing Spacetime and the Early Universe}\label{sect1}

It is widely believed that our Universe went through an exponentially accelerated expansion at the very early time. This process, known as ``inflation'' \cite{gut}, explains the flatness problem (why is the spatial curvature so close to zero), the horizon problem (why is the CMB temperature isotropic in all directions that we look, despite those regions having no causal contact in a standard Big Bang cosmology without inflation), and the monopole problem (why there is no magnetic monopole). Inflation explains these by essentially ``washing away'' all irregularities. Nevertheless, it has been argued that inflation by itself does not explain the arrow of time \cite{page1, page2, hp, 0205058v2, 0505037}, as usually inflation itself requires special initial conditions to occur. For example, the simplest models with single inflaton field requires a ``slow-roll'' condition (see the discussions in \cite{0911.0693, 0711.1656v2}). See, however, \cite{brand, 1601.01918}.



In addition to scalar fields, the inflationary epoch of the Universe may also involve spacetime shearing effects\footnote{Not to be confused with ``cosmic shear'',  which is the distortion of images of distant galaxies due to weak gravitational lensing by the large scale structure in the Universe \cite{1411.0115}.}. Although the presence of initial inhomogeneities and anisotropies, if any, will likely be washed away by inflation (if inflation can start), since not much is known about inflation, we cannot yet rule out models in which cosmic shears remain after inflation (see the discussion involving ``vector inflation'' in \cite{1509.09166}; also shear viscous effects can arise in warm inflation \cite{1106.0701}), although the measurement of an almost-isotropic distribution of the temperature of the cosmic microwave background radiation suggests that they are negligible in the present epoch \cite{2013a}. In fact, from a purely mathematical perspective it can be proved that at least within the homogeneous but anisotropic Bianchi I models with regular matter content, a shear term dominates the primordial evolution, which is well approximated by the Kasner solution, subsequently becoming negligibly small at late times \cite{kasner1,kasner2}. (In the presence of a cosmological constant, Bianchi models lack ``primordial anisotropic hair'' \cite{wald} . Such a ``cosmic no-hair theorem'' can be circumvented, however, in the presence of vector fields \cite{ford, 0802.2068, 0805.4229, 0805.4265, 0902.2833}, for example). However, we stress that in general it is not enough to observe the isotropy of one physical quantity, like the temperature of the cosmic microwave background radiation, for claiming that \emph{all} the other cosmological parameters should also be isotropic as well \cite{lim1,lim2}. For example, statistical anisotropy can appear in the bispectrum of curvature perturbation even if it does not appear in the power spectrum \cite{0805.4265}. In addition, large-scale asymmetries and alignments of astrophysical filaments along a preferred spatial direction have been observed \cite{asy1,asy2,asy3,asy4,asy5}, e.g., the so-called ``axis of evil'' \cite{0502237}. Nevertheless, there is as yet no consensus as to how much of these effects are due to systematical or contaminative errors in observation or in data analysis \cite{1708.06139}.

An analytical and exact solution of the Einstein's field equations of general relativity entirely written in terms of elementary functions describing both nontrivial shearing and expansion effects supported by a massless scalar field in both open and closed topologies, has been investigated in \cite{ref3,ref4,ref5,ref6,ref7} (see also page 261 of \cite{exact} for a summary). Therefore, it is important to discuss the physical viability of this class of cosmological spacetimes as a realistic model of our Universe, at least at the early times. One such check is by investigating their thermodynamical properties. In the cosmological context, one of the theoretical thermodynamical constraint is formulated as the ``cosmological holographic principle'', according to which the amount of matter entropy inside the region bounded by the dynamical apparent horizon must not be larger than the area of the horizon itself\footnote{In the commonly used Planck units, in which $\hbar=G=c=k_B=1$, with $\hbar$ the reduced Planck constant, $G$ the Newton's constant, $c$ the speed of light, and $k_B$ the Boltzmann's constant, the precise statement would be $S\leqslant A/4$ with $S$ denoting the entropy and $A$ the area. For our purpose, it is enough to consider $S \lesssim A$.
In this work, however, we do not employ the Planck units.} \cite{bousso}. Another requirement is that a physical model should satisfy the second law of thermodynamics, which requires a non-decreasing entropy in time. Our goal is to clarify which, of any, of the two topologies is favoured by these two requirements and possibly set an upper bound for the shear at early times with respect to the other cosmological parameters, namely the Hubble function and the matter parameter. In fact, we want to extend our way of thinking, which has already been proven as a valid tool for constraining the strength of spatial inhomogeneities for the spherically symmetric Stephani universe (in which pressure is a function of both space and time), to the inflationary epoch. In that case, the second law could be recast as an independent and complementary estimate of the present day \lq\lq acceleration" of the Universe without relying on astrophysical measurements \cite{stephani}. Morever, it should be emphasized that, the cosmological holographic principle is a powerful tool in testing dark energy models in late-time cosmology \cite{hol1,hol2,hol3}, inhomogeneous cosmological models such as the Lema\^itre-Tolman-Bondi model (in which density is a function of space and time) \cite{hol4}, the number of spatial dimensions of the Universe \cite{hol5}, and the cosmic microwave background signatures \cite{hol6,hol7}, just to mention a few applications.

Furthermore, while primordial quasars and galaxies containing a supermassive black hole have been observed even at redshift $z\sim 10$ \cite{form1,form2},
perturbation theory applied to a homogeneous and isotropic Friedmann universe and standard accretion mechanisms cannot account for their existence \cite{form3,form4}. Thus, it has been argued that inhomogeneous shearing spacetimes supported by a massless scalar field may provide a valid framework for their description without the need to invoke any quantum modification to general relativity \cite{form5,form6}.

The cosmological solutions we will study in this paper are algebraically Petrov type D, unlike the Friedmann metric which is of Petrov type O (conformally flat with only Ricci curvature). In other words, some Weyl curvature affects the evolution of the matter content and of the whole Universe. Therefore, we have in hands an exact framework for testing the {Weyl curvature conjecture}, which states that the gravitational entropy in a non-stationary spacetime should be proportional to the square of the Weyl tensor, which consequently must grow during the time evolution \cite{wcc1,wcc2,wcc3,wcc4}, for complementing the previous literature studies in homogeneous cosmologies \cite{wcc5}, and black hole physics \cite{wcc6,wcc7}, which include black rings \cite{wcc8}.

In this manuscript we will provide a more transparent physical interpretation of a class of mathematical solutions of the Einstein's equations of General Relativity found by Leibovitz-Lake-Van den Bergh-Wils-Collins-Lang-Maharaj \cite{ref3,ref4,ref5,ref6,ref7} by proposing a novel set of conditions on the free parameters of their model. First and foremost, we require that the energy density of the cosmic fluid must be positive. Adopting a modern language, the cosmic fluid is interpreted as a so-called ``chameleon field'' \cite{amanda1, amanda2} because its equation of state parameter is energy-dependent, and as a massless scalar field following the canonical formalism. After that, the evolution of these spacetimes is further constrained in light of the cosmological holographic principle and the second law of thermodynamics, complementing our previous study of the shear-free and conformally flat Stephani model \cite{stephani}. We will also comment on the consequences on the sign of the deceleration parameter, which will be shown to be negative after imposing those requirements,  and before relying on any astrophysical datasets.

\section{Some exact cosmological shearing solutions with a massless scalar field} \label{sect2}

In this section we will introduce the cosmological models that we want to investigate in light of the cosmological holographic principle, and of the second law of thermodynamics. Firstly, we will derive some constraints for their free parameters by requiring that the energy density of the cosmic fluid must be non-negative, and then we will compute the kinematical variables characterizing the evolution of this spacetime.

In a spherical coordinate system $x^\mu$=($t$, $r$, $\theta$, $\phi$), and adopting the Lorentzian segnature $(-,+,+,+)$, the spacetime metric tensor

\beq
\label{eq1}
\d s^2=g_{\mu\nu}\d x^\mu \d x^\nu=-\left(\frac{c r}{2 l}\right)^2 \d t^2 +\frac{\d r^2}{\epsilon +C r^2}+r^2  \left[\frac{\epsilon}{2} +h(t) \right] (\d\theta^2 +\sin^2 \theta \d \phi^2),
\eeq
with:
\begin{eqnarray}
\label{eqh1}
h(t)&=& A \sin (ct/l) +B \cos (ct/l) \qquad {\rm if} \qquad \epsilon=-1 \,, \\
\label{eqh2}
h(t)&=& -\left(\frac{ct}{2l} \right)^2+ \frac {2Act} {l} +B  \qquad {\rm if} \qquad \epsilon=0 \,, \\
\label{eqh3}
h(t)&=& A e^{ct/l} +B e^{-ct/l} \qquad {\rm if} \qquad \epsilon=1,
\end{eqnarray}
is an exact solution of the Einstein's field equations of general relativity, $G_{\mu\nu}=({8 \pi G}/{c^4})T_{\mu\nu}$, for a perfect fluid whose equation of state relating pressure and energy density is\footnote{The reader can find these information summarized on page 261 of \cite{exact}.} \cite{ref3,ref4,ref5,ref6,ref7}
\beq
\label{eos}
p=c^2\rho+\frac{3 c^4 C}{4 \pi G}\,.
\eeq
The former two account for an open topology of the universe, while the latter for a closed one.
The stress-energy tensor for the matter content is $T_{\mu\nu}=(\rho+p/c^2)u_\mu u_\nu+(p/c^2) g_{\mu\nu}$, in which we have introduced the observer four-velocity $u^\mu=dx^\mu/d\tau={c \delta_t^\mu}/{\sqrt{-g_{tt}}}$, $u_\mu u^\mu=-c^2$. Moreover, the constants $A$, $B$ and $C$ are the free parameters of the model which are not constrained by the field equations, and $\epsilon$ accounts for the topology of the universe. Note that $A$, $B$, $\epsilon$ and $ct/l$ are dimensionless quantities, while [C]=L$^{-2}$. In addition, $l$ is a reference length scale which has been introduced for dimensional purposes, that from now on we will assume to be unity without loss of generality because it can be re-absorbed into the time coordinate as a rescaling factor. The equation of state of the cosmic fluid can be re-written in the form
\beq
\label{pressure}
p= w(\rho) \rho\,, \qquad w(\rho)=c^2+\frac{3 c^4 C}{4 \pi G \rho},
\eeq
in which the constant equation of state parameter adopted in the standard cosmological modeling has been promoted to an energy-dependent {\it chameleon field} \cite{amanda1,amanda2}, so named because the range of the force mediated by the scalar particle becomes small in regions of high density, but shows its effect at large cosmic distances. In particular, the comic fluid reduces to an ideal fluid with its energy density and pressure being directly proportional to each other in the high energy regime, exhibiting the same {\it asymptotic freedom} which characterizes the bag model of quark-gluon plasma with the constant $C$ playing the role of vacuum energy \cite{gross1,gross2}. This also leads to a similar notion of ``bag energy'' as one finds in the contexts of quark physics, as will be discussed later. The chameleon properties of the cosmic fluid are suppressed in the limit $C \to 0$. However, in this case the spacetime metric (\ref{eq1}) would be ill-defined both in the cases $\epsilon=-1$ and $\epsilon=0$ because of the unphysical Lorentz signature in its $g_{rr}$ component;  while for the choice $\epsilon=1$ such parameter would not play any role because it can be re-absorbed through a re-scaling of the radial coordinate $r$.

The adiabatic speed of sound within the fluid is $c_s=\sqrt{\frac{\partial p}{\partial \rho}}=c$ which is the same as a stiff fluid. Taking into account the canonical equations of the ``fluid-scalar field correspondence'' \cite{corr1,corr2}, the Einstein's equations for a scalar field $\Phi$ minimally coupled to gravity can be derived by applying a variational principle to the total Lagrangian
\beq
{\mathcal L}={\mathcal L}_{\rm EH}+{\mathcal L}_{\rm m}\,,
\eeq
where ${\mathcal L}_{\rm EH}$ is the Einstein-Hilbert part, and the matter contribution can be written in terms of the kinetic and potential energy of the scalar field
\beq
{\mathcal L}_{\rm m}=-\frac{1}{2} g^{\mu\nu}\partial_\mu \Phi \partial_\nu \Phi +V(\Phi)\,.
\eeq
In fact, the canonical equations, when the gradient of the scalar field is timelike, allow us to express the energy and pressure of a perfect fluid in terms of the kinetic and potential energy of the underlying scalar field as \cite{qqq3,kolb,kras}
\beq
 c^2 \rho=\frac{\Phi_{;\mu}\Phi^{;\mu}}{2}+V(\Phi)\,, \qquad  p=\frac{\Phi_{;\mu}\Phi^{;\mu}}{2}-V(\Phi)\,.
\eeq
Thus, the equation of state (\ref{eos}) can be re-interpreted as describing a free inhomogeneous scalar field inside a constant potential $V(\Phi)=-{3 c^4 C}/{8 \pi G}$. As is well known, an additive constant entering the Lagrangian can be re-absorbed, shifting the zero energy level of the system, and does not affect the dynamical evolution of the scalar field, which is still governed by the free Euler-Lagrange equation $\Box \Phi:=g^{\mu\nu} \nabla_\mu \partial_\nu \Phi=0$. Therefore, the spacetime under investigation is permeated by a fluid which behaves {\it effectively} as stiff matter, and consequently as a massless scalar field, from the hydrodynamic point of view. The relation (\ref{eos}) is named {\it stiffened equation of state} and constitutes a simplified version of the Gr\"uneisen model, in which the constant $C$ takes into account the deviations which occur at high pressure which are likely to be realized in the early Universe \cite{stiffe1,stiffe2,stiffe3}. 
Massless scalar fields (or equivalently stiff fluids) have already been adopted in the modeling of the early Universe. For example, they are a basic assumption in the formulation of the Belinskii-Khalatnikov-Lifschitz (BKL) locality conjecture for studying the Big Bang spacelike singularity \cite{bkl1,bkl2,bkl3}. Furthermore, energy exchanges with a massless scalar field may cause the accretion of primordial black holes \cite{carr}, and more generally  a stiff matter dominated era occurs both in the Zel'dovich model of a primordial universe constituted by a cold gas of baryons \cite{zel},  and when the cosmic fluid is represented by a relativistic self-gravitating Bose-Einstein condensate  \cite{sss}.    

 A peculiar property of the universe modeled by (\ref{eq1}) is that the cosmic expansion affects only its angular part, but not the radial one. The physical interpretation is that a measured non-zero gravitational redshift would imply that the light rays coming from galaxies are traveling along non-radial orbits. This would be a consequence of the particular gravitational field shaping this spacetime, and the purpose of our manuscript is to investigate its signatures on the formation of some primordial structures whose existence cannot be accounted for within the standard model of cosmology.

We can claim that the metrics (\ref{eq1}) with  the function $h(t)$ defined by (\ref{eqh1})-(\ref{eqh2})-(\ref{eqh3}) are spatially inhomogeneous by considering the $r$-dependence affecting the Ricci scalar (which is a curvature invariant independent of the system of coordinates \cite{mcc}):
\beq
\label{RS}
R= -2 \frac{\epsilon+{\mathcal R}+6 C r^2 (\epsilon+2h(t))^2}{r^2(\epsilon+2h(t))^2},
\eeq
in which
\begin{eqnarray}
\label{curv}
{\mathcal R}&=& 4(A^2+B^2) \qquad {\rm for} \qquad \epsilon=-1 \,, \\
{\mathcal R}&=& 4(4A^2+B) \qquad {\rm for} \qquad \epsilon=0 \,, \nonumber\\
{\mathcal R}&=& -16AB  \qquad {\rm for} \qquad \epsilon=1 \,. \nonumber
\end{eqnarray}
We can compute the energy density of the cosmic fluid by inserting the equation of state (\ref{eos}) into the trace of the Einstein's field equations:
\beq
\label{rho}
\rho= - \frac{c^2 (R + 18 C)}{16\pi G} \,,
\eeq
where the Ricci scalar $R$ has been given in (\ref{RS}). A well-defined (non-negative) energy density requires
\beq
\label{conC}
C \leqslant \frac{\epsilon+{\mathcal R}}{3 r^2 (\epsilon+2 h(t))^2}.
\eeq
Furthermore, the energy density exhibits the two limiting behaviors
\beq
\label{rhoinf}
\rho_0:=\lim_{r \to 0} \rho = \infty \cdot {\rm sgn} (\epsilon + {\mathcal R}) \,, \qquad \rho_{\infty}  :=\lim_{r \to +\infty} \rho = -\frac{3 C c^2}{8\pi G},
\eeq
at the center of the configuration and in the far field limit, respectively, where sgn denotes the sign function. The latter expression shows that a non-negative energy density requires $C \leqslant 0$, which opens up the possibility of having a negative pressure from (\ref{eos}) in certain spatial regions and/or at certain time intervals mimicking a cosmological constant term. In fact, in this asymptotic regime the effective equation of state of the cosmic fluid reduces to $p_{\infty}=-c^2 \rho_{\infty}$. Moreover, $\rho_0 \geqslant 0$ implies $\epsilon + {\mathcal R} \geqslant 0$, which in turn is equivalent to the following constraints between the free parameters in the three topologies:
\begin{eqnarray}
\label{hyperbola}
4(A^2 +B^2)-1& \geqslant & 0 \qquad {\rm for} \qquad \epsilon=-1 \,, \\
4 (4 A^2 +B)& \geqslant & 0\qquad {\rm for} \qquad \epsilon=0 \,, \nonumber\\
1-16AB& \geqslant &0  \qquad {\rm for} \qquad \epsilon=1 \,. \nonumber
\end{eqnarray}
Therefore, the two parameters $A$ and $B$ live in a phase region bounded by a circumference of a circle, a parabola, and a hyperbola, respectively. Defining the Big Bang time $t_{\rm BB}$  as the time at which the energy density diverges, we can conclude that it is given implicitly by the condition $\epsilon +2 h(t_{\rm BB})=0$. Therefore, in our model of the Universe the time at which the initial singularity occurs is affected both by the topology and by the parameters $A$ and $B$, but not $C$. A qualitative difference with the more popular Lema\^itre-Tolman-Bondi is that the Big Bang time is not space-dependent \cite{bangltb1,bangltb2}.

The kinematical variables characterizing the spacetime described  by metric (\ref{eq1}) can be computed following  \cite{kasner1}. The Hubble function is given by
\beq
\label{HH}
H:=\frac{u^{\mu}{}_{;\mu}}{3}= \frac{4 \dot h(t)}{3(\epsilon +2 h(t))r},
\eeq
which diverges and vanishes for small and large $r$ respectively, and is monotonically decreasing in between. In this formula an over dot denotes a derivative with respect to the coordinate time. Interestingly, in this model the Hubble function is inhomogeneous allowing us to complement our previous analysis based on the Stephani model, which instead exhibits a homogeneous rate of expansion \cite{stephani,homH}.  We note that an exponential evolution of the scale factor can imply an almost-constant Hubble rate. Thus, taking into account also that the matter source is a scalar field, and that there are some spacetime shearing effects, our model with the choice (\ref{eqh3}) is suited for describing the early inflationary stages of the universe \cite{corr1}. 
	
\begin{figure}
	\begin{center}
		$
		\begin{array}{cc}
		{\includegraphics[scale=0.5, angle=0]{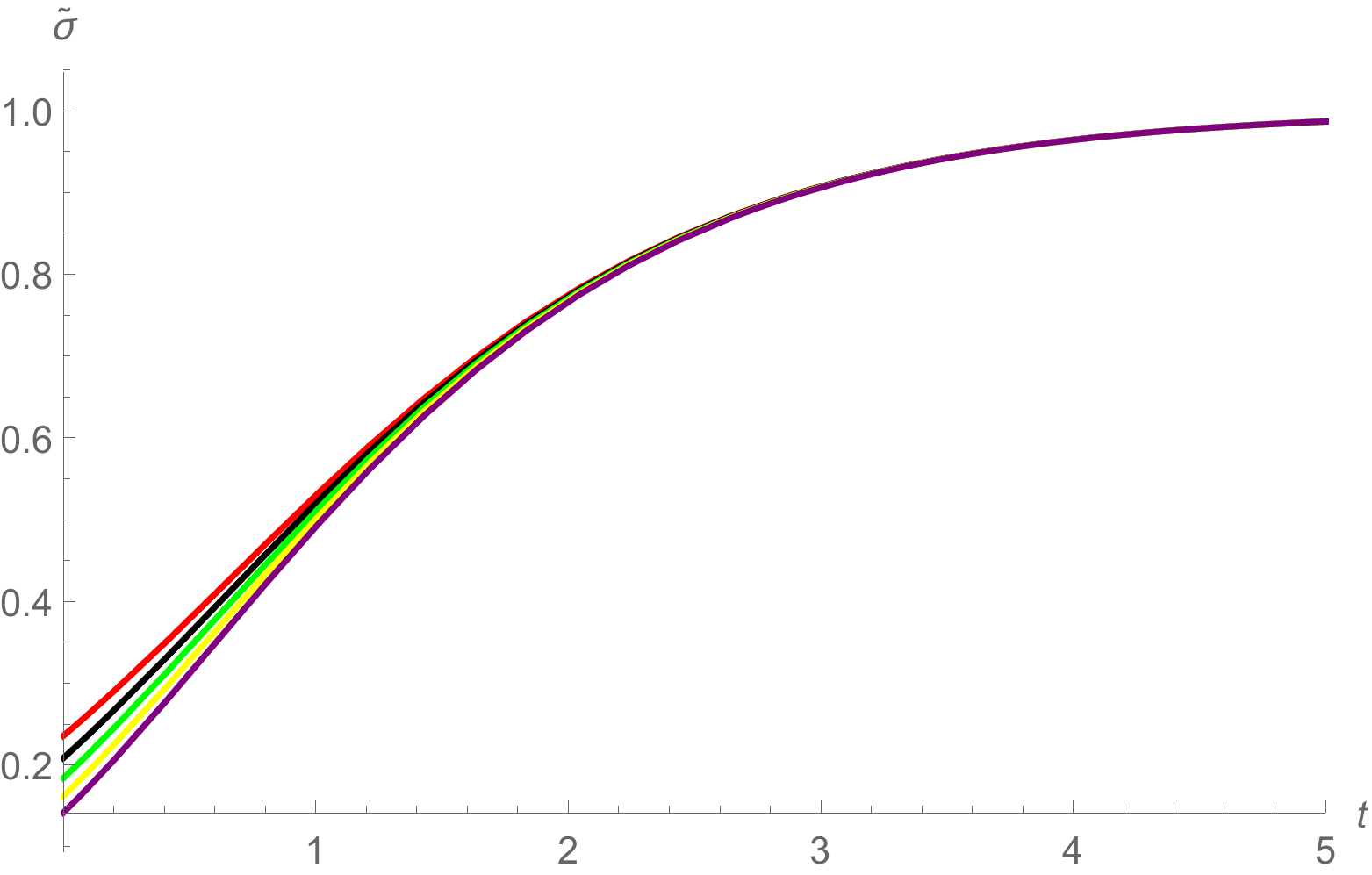}}
		\hspace{3mm}
		{\includegraphics[scale=0.5, angle=0]{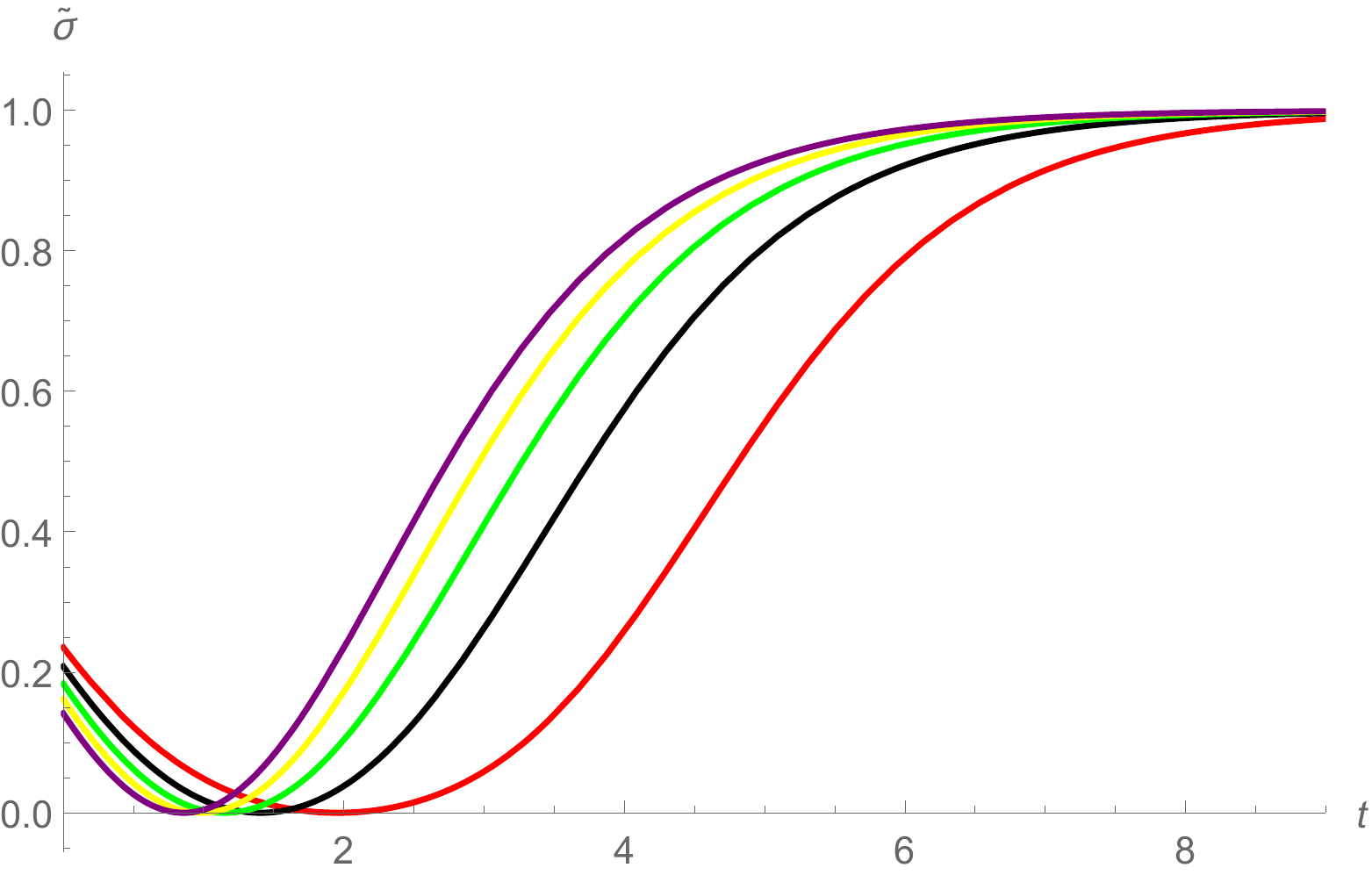}}\\
		(a) \hspace{90mm} (b) 
		\end{array}
		$
	\end{center}
	\caption{The figure depicts  the time evolution of the quantity $\tilde \sigma:= 3r^2 \sigma^2$ computed from (\ref{shear}) for the universe (\ref{eqh3}) at the location $r=1$, and in units such that $c=1$. In panel (a) we choose A=0.50 and B=0.01 (red line),  B=0.03 (black line),  B=0.05 (green line),  B=0.07 (yellow line),  B=0.09 (purple line); in panel (b) we choose B=0.50 and A=0.01 (red line),  A=0.03 (black line),  A=0.05 (green line),  A=0.07 (yellow line),  A=0.09 (purple line). We remark that our choices for the numerical values of the free parameters are consistent with (\ref{hyperbola}). In Sect. \ref{law} we will show that our model is physically acceptable in light of the second law of thermodynamics only at times $t>\frac{1}{2c}\ln\frac{B}{A}$, for which we can see that the shear would be monotonically increasing.
	}
	\label{fig1}
\end{figure}

	In fact, the  shear tensor reads
\beq
\sigma_{ij}={\rm diag}\Big[ -\frac{4 \dot h(t)}{3(C r^2 +\epsilon)(\epsilon+2 h(t)) r}, \, \frac{\dot h(t) r}{3}, \, \frac{\dot h(t) r \sin^2 \theta}{3}\Big]\,, \qquad i,j\,=\,r,\theta,\phi\,,
\eeq
which implies 
\beq
\label{shear}
\sigma^2 := \frac{1}{2} \sigma_{ij} \sigma^{ij}= 
\frac{4 \dot h(t)^2}{3 (\epsilon +2 h(t))^2  r^2}=\frac{3H^2}{4}\,.
\eeq
Thus, the shear displays a more severe divergence towards the center of the configuration, and it asymptotes to zero faster at spatial infinity than the Hubble function. Moreover, the spacetime shearing effects are bounded by the rate of expansion of the Universe $\sigma/H<1$ in agreement with standard CMB physics \cite{borgani}. 
 We display in Fig. \ref{fig1} the time evolution of the quantity $\tilde \sigma:= 3r^2 \sigma^2$ for the universe (\ref{eqh3}) at the location $r=1$, and in units such that $c=1$. In panel (a) we choose A=0.50 and B=0.01 (red line),  B=0.03 (black line),  B=0.05 (green line),  B=0.07 (yellow line),  B=0.09 (purple line); in panel (b) we choose B=0.50 and A=0.01 (red line),  A=0.03 (black line),  A=0.05 (green line),  A=0.07 (yellow line),  A=0.09 (purple line). We remark that our choices for the numerical values of the free parameters are consistent with (\ref{hyperbola}). In Sec. \ref{law} we will show that our model is physically acceptable in light of the second law of thermodynamics only at times $t>\frac{1}{2c}\ln\frac{B}{A}$, for which the shear would be monotonically increasing.

This class of metrics also exhibits a non-trivial acceleration vector
\beq
\dot u^\mu:=u^\nu \nabla_\nu u^\mu= \frac{Cr^2+\epsilon}{r}c^2 \delta^\mu_r\,.
\eeq
Then, the generalized Friedmann equation (which is the mixed-rank time-time component of the Einstein's equations) allows us to compute the spatial curvature as \cite{kasner1}:
\beq
^3R=\frac{16 \pi G \rho}{c^2}-6 H^2+2\sigma^2= -R-18C-\frac{9H^2}{2} =2 \left[  \frac{\epsilon + {\mathcal R} -4 \dot h(t)^2}{(\epsilon +2 h(t))^2 r^2}    -3C \right]   \,,
\eeq
where in the last step we used (\ref{shear}) and (\ref{rho}). The first equality means that in this model of the Universe, unlike the Friedmann cosmology, the evolution of the Hubble function is affected not only by the energy density permeating the space, but also by a certain linear combination of the invariant shear and of the spatial curvature. The Stephani universe  exhibits a similar behavior because the evolution of the Hubble function is affected not only by the abundance of regular matter within the spacetime, but also by the strength of spatial inhomogeneities which plays the role of an effective mass-energy parameter as we discussed in our previous work \cite{stephani}. However, an important difference is that the former spacetime is shear-free. Adopting the standard terminology, we can introduce the matter density parameter 
\beq
\Omega_{\rm m}=\frac{8\pi G}{3H^2} \rho= \frac{3 c^2}{  16 \dot h(t)^2} [\epsilon + {\mathcal R} -3 C (\epsilon +2 h(t))^2 r^2 ]\,,
\eeq
using (\ref{rho}) and (\ref{HH}). We may note that the matter density parameter is regular even at $r=0$ because the divergence in the Hubble function has canceled the divergence in the energy density.

Unlike the Friedmann spacetime, which is conformally flat, the model under consideration (\ref{eq1}) displays a non-trivial Weyl curvature tensor $C_{\mu\nu\rho\sigma}$  because it is of the algebraic Petrov type D. We quantify the strength of the Weyl curvature applying the Newman-Penrose formalism \cite{exact,advgr,NPC}. Let
\begin{eqnarray}
\label{tetrad}
l^a &=& \frac{\sqrt{2} cr}{4} \d t -\frac{\d r}{\sqrt{2(Cr^2 +\epsilon)}} \,, \\
n^a &=& \frac{\sqrt{2} cr}{4} \d t +\frac{\d r}{\sqrt{2(Cr^2 +\epsilon)}}  \,, \nonumber\\
m^a &=& \frac{r \sqrt{\epsilon +2 h(t)}}{2} (\d\theta +i \sin \theta \d\phi)\,, \qquad i^2=-1, \nonumber
\end{eqnarray}
be a null tetrad such that
\beq
l_a l^a=n_a n^a=m_a m^a =\bar m_a \bar m^a=0\,, \qquad -l_a n^a=1=m_a \bar m^a,
\eeq
where an over bar stands for complex conjugation, in terms of which the metric (\ref{eq1}) can be written in the form $\d s^2=-2l_{(a} n_{b)}+2 m_{(a} \bar m_{b)}$, where round parentheses denote symmetrization.  The coframe (\ref{tetrad}) provides the canonical form of the Newman-Penrose scalars related to the Weyl curvature tensor  because $\Psi_0=\Psi_1=\Psi_3=\Psi_4=0$,  and
\beq
\label{weylcurv}
\Psi_2=-\frac{{\mathcal R} +\epsilon}{3 r^2 (2 h(t)+\epsilon)^2}\,,
\eeq
where ${\mathcal R}$ can be obtained from (\ref{curv}). Thus,  the quantity $\Psi_2$ contains all the information we need about the Weyl curvature. We note that $\Psi_2$ accounts for a Coulomb-like gravitational potential \cite{peeling} and it is related to the ``electric'' ($E_{\mu\nu}$) and ``magnetic'' ($B_{\mu\nu}$) Weyl components through (since it is purely real in our case) \cite{exact}:
\beq
\Psi_2^2= \frac{E_{\mu\nu}E^{\mu\nu}-B_{\mu\nu}B^{\mu\nu}}{6} \,.
\eeq

\begin{figure}
		\begin{center}
			$
			\begin{array}{cc}
			{\includegraphics[scale=0.5, angle=0]{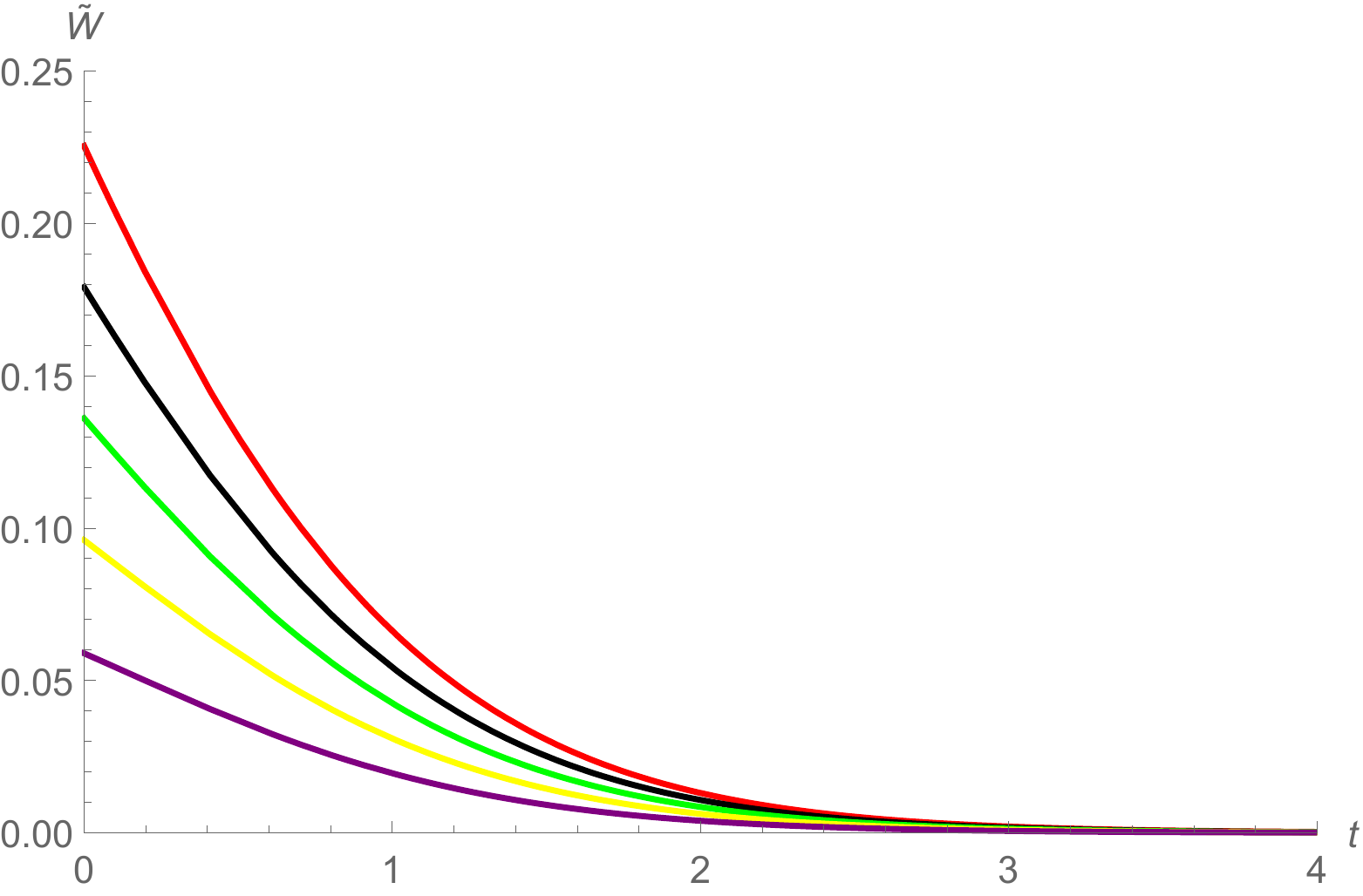}}
			\hspace{3mm}
			{\includegraphics[scale=0.5, angle=0]{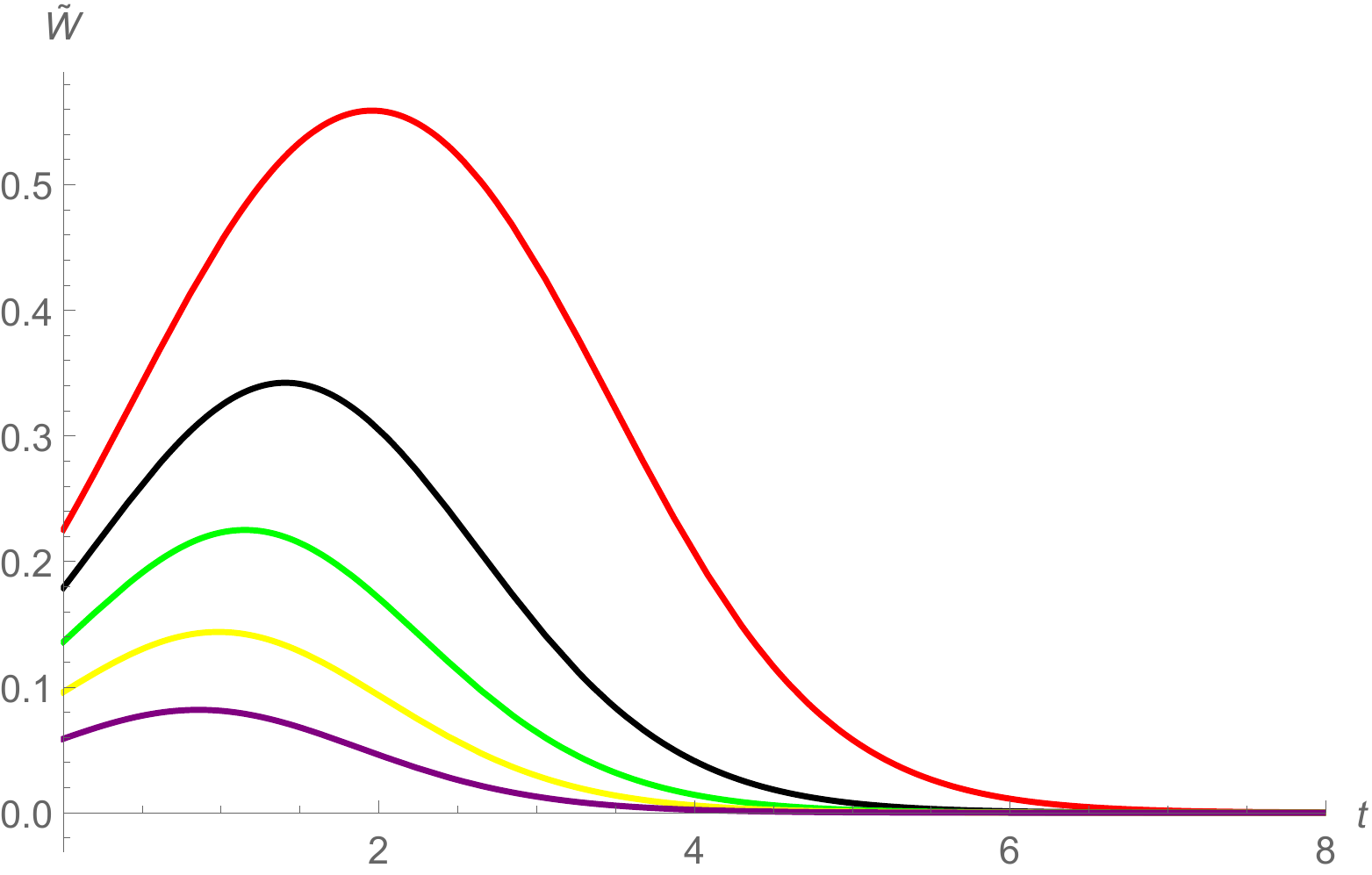}}\\
			(a) \hspace{90mm} (b) 
			\end{array}
			$
		\end{center}
		\caption{The figure depicts  the time evolution of the quantity $\tilde W:= 3r^2 |\Psi_2|$ computed from (\ref{weylcurv}) for the universe (\ref{eqh3}) at the location $r=1$, and in units such that $c=1$. In panel (a) we choose A=0.50 and B=0.01 (red line),  B=0.03 (black line),  B=0.05 (green line),  B=0.07 (yellow line),  B=0.09 (purple line); in panel (b) we choose B=0.50 and A=0.01 (red line),  A=0.03 (black line),  A=0.05 (green line),  A=0.07 (yellow line),  A=0.09 (purple line). We remark that our choices for the numerical values of the free parameters are consistent with (\ref{hyperbola}). In Sec. \ref{law} we will show that our model is physically acceptable in light of the second law of thermodynamics only at times $t>\frac{1}{2c}\ln\frac{B}{A}$, for which we can see that the Weyl curvature would be monotonically decreasing.  Comparing with Fig. \ref{fig1} we can understand that in this model of the universe the shear is increasing when the Weyl curvature is decresing and viceversa.
		}
		\label{fig2}
	\end{figure}

 We display in Fig. \ref{fig2} the time evolution of the quantity $\tilde W:= 3r^2 |\Psi_2|$ for the universe (\ref{eqh3}) at the location $r=1$, and in units such that $c=1$. In panel (a) we choose A=0.50 and B=0.01 (red line),  B=0.03 (black line),  B=0.05 (green line),  B=0.07 (yellow line),  B=0.09 (purple line); in panel (b) we choose B=0.50 and A=0.01 (red line),  A=0.03 (black line),  A=0.05 (green line),  A=0.07 (yellow line),  A=0.09 (purple line). We remark that our choices for the numerical values of the free parameters are consistent with (\ref{hyperbola}). In Sec. \ref{law} we will show that our model is physically acceptable in light of the second law of thermodynamics only at times $t>\frac{1}{2c}\ln\frac{B}{A}$, for which the Weyl curvature would be monotonically decreasing. Comparing with Fig. \ref{fig1} we can understand that in this model of the universe the shear is increasing when the Weyl curvature is decresing and viceversa.
	
Finally, the covariant deceleration parameter in this class of metrics is given by \cite{kasner1}:
\beq
q =\frac{1}{H^2}\left(2  \sigma^2 -\frac{\frac{\tilde \nabla_\mu \dot u^\mu }{c} + \frac{\dot u_\mu \dot u^\mu }{c^2}}{3}   \right) + \frac{\Omega_{\rm m}}{2} \left(1+\frac{3 \omega}{c^2} \right),
\eeq
in which we have introduced the notation $\omega=p/\rho$ for the equation of state parameter, and 
\beq
\tilde \nabla_\mu \dot u^\mu=h^{\mu}{}_{\nu} h^{\tau}{}_{\mu} \nabla_\tau \dot u^\nu=h^{\tau}{}_{\nu} \nabla_\tau \dot u^\nu\,,
\eeq 
for the fully orthogonally projected covariant derivative, where $h_{\mu\nu}=g_{\mu\nu}+u_\mu u_\nu $ is the spatial metric. Therefore, the deceleration parameter can be written explicitly as a function of $\epsilon$ and $h(t)$, and the derivative of $h(t)$:
\beq
q=3 \left[ 1-\frac {c^2 (\epsilon + 2 h(t))(\epsilon^2 +2 \epsilon h(t) -1)}{4 \dot h(t)^2}   \right]\,.
\eeq
We note that the parameter $C$ does not play any direct role. Furthermore, the deceleration parameter is spatially homogeneous, contrary to the case of the Stephani universe \cite{homH}.

\section{A thermodynamical estimate of the cosmological parameters} \label{sect3}

We are now ready to investigate how the cosmological holographic principle and the second law of thermodynamics can provide a set of constraints between various cosmological parameters (deceleration parameter, expansion, shear, matter-energy abundance, and curvature strength) complementary and independent to those which may come from astrophysical observations.

We start by recalling that the location of the dynamical apparent horizon follows from the condition $|| \nabla \tilde r||^2=0$, where $\tilde r= r \cdot \sqrt{\frac{\epsilon +2 h(t)}{2}}$ is the areal radius \cite{ash}. Explicitly we must solve the algebraic equation
\beq
(C r^2 +\epsilon) c^2 (\epsilon +2 h(t))^2 -4 \dot h(t)^2=0 \,,
\eeq
which admits a non-complex solution only for the closed topology $\epsilon=1$ (taking into account that $C<0$ from (\ref{rhoinf})),  and as long as
\beq
\label{conhor}
r>\sqrt{-\frac{1}{C}}\,.
\eeq
In this latter case two mathematical solutions
\beq
r_{1,2}= \pm \frac{1}{c (\epsilon +2 h(t))} \cdot \sqrt{\frac{4 \dot h(t)^2 -c^2 \epsilon (\epsilon +2 h(t))^2}{C}} 
\eeq
can be found, of which only the positive root is of physical interest. Therefore, the dynamical apparent horizon is located at:
\begin{eqnarray}
\label{apparenthor}
r_{\rm AH} &=& \sqrt{\frac{1}{C}\left[\frac{2 \dot h(t)^2}{c^2 (\epsilon + 2h(t))}-\frac{\epsilon (\epsilon + 2h(t))}{2}   \right]} \\
&=& \sqrt{-\frac{16 AB +4 e^{tc} + 4 e^{-tc}+1 }{2C(2 A e^{tc}  +2 B e^{-tc} +1)}}=\sqrt{\frac{ {\mathcal R} -4 h(t) -\epsilon}{ 2C (\epsilon + 2h(t))}} \,, \nonumber
\end{eqnarray}
where in the last step we have specialized the result to the topology $\epsilon=1$. The existence of the square root can be easily guaranteed by restricting both $A$ and $B$ to be positive. In this case, taking into account (\ref{hyperbola}), we obtain a further constraint  
\beq
A \cdot B \leqslant \frac{1}{16} \,.
\eeq
For the topology with $\epsilon=1$, the deceleration parameter can be re-written as:
\beq
q=-\frac{3 ( - {\mathcal R}/2 + h(t))c^2}{2 \dot h(t)^2}=-\frac{3 (8AB +A e^{tc} + B e^{-tc})}{2(A e^{tc} - B e^{-tc})^2}\,,
\eeq
which is automatically negative if both $A$ and $B$ are positive. Moreover,  the choices of positive $A$ and $B$ -- if taken at face value -- make the model with $\epsilon=1$ come without a big bang singularity because $\epsilon +2 h(t) \neq 0$ $\forall t \in \mathbb{R}$, and further requiring a negative $C$ together with the condition (\ref{conhor}) they preserve the Lorentzian signature and the causality structure of the spacetime (\ref{eq1}). Nevertheless, as we shall see the model is only valid after some time $t>0$.

\subsection{Cosmological Constraints From the Cosmological Holographic Principle}

According to the cosmological holographic principle, the matter entropy, $S_{\rm m}$, inside the region bounded by the dynamical apparent horizon should be smaller than the area $A_{\rm AH}$ of this spacetime region \cite{bousso}. In the case of the universe (\ref{eq1}), for the topology for which a dynamical apparent horizon indeed exists, its area  is \cite{toolkit}
\beq
A_{\rm AH}= 4\pi r_{\rm AH}^2=\frac{2 \pi ({\mathcal R} -4h(t) -\epsilon)}{C (\epsilon + 2h(t))}=     -2\pi\frac{4(4AB+A e^{tc}+B e^{-tc})+1}{C(2A e^{tc}+2B e^{-tc}+1)}\,.
\eeq
The entropy of the matter content inside the spacetime region bounded by the dynamical apparent horizon is \cite{stephani}:
\beq
\label{cubic}
S_{\rm m}= {\tilde \alpha} \, r_{\rm AH}^3=  {\tilde \alpha}  \left[\frac{ {\mathcal R} -4 h(t) -\epsilon}{ 2C (\epsilon + 2h(t))} \right]^{3/2}  = {\tilde \alpha} \left[-\frac{4(4AB+A e^{tc}+B e^{-tc})+1}{2C(2A e^{tc}+2B e^{-tc}+1)}\right]^{3/2}\,.
\eeq
The constant
\beq
{\tilde \alpha}=\frac{4 k_B ^4}{135}\left(\frac{\pi T c^2}{\hbar}  \right)^s (1 + z_e)^s\,, \qquad s=6,
\eeq
summarizes all the information about the cosmic fluid. In more details,  $k_B$ is the Boltzmann constant that enters the Boltzmann law of black body radiation, $\hbar$ is the reduced Planck constant, $T$ is the temperature of the cosmic fluid, and $z_e$ is the redshift at the decoupling era. The power factor $s=3(1+w)$,  which accounts for the stretching of wavelengths in an expanding Universe (Hubble law),  has been computed for the equation of state parameter $w=1$ that characterizes a stiff fluid. It is important not to confuse the factor $s$ which depends on the type of the matter content inside the region bounded by the dynamical apparent horizon, and the geometrical factor 3 in the first equality of (\ref{cubic}), which instead is needed for computing the volume of this region.

Mathematically, the condition which follows from the cosmological holographic principle is: 
\beq
\frac{S_{\rm m}}{A_{\rm AH}} < \frac{1}{4 L_p^2}=\frac{c^3}{4 G \hbar} \qquad  \Rightarrow \qquad \frac{\alpha }{4 \pi}r_{\rm AH}<1 \,,
\eeq
with $\alpha=4{\tilde \alpha}L_p^2  $, $L_p$ being the Planck length.
Observing that the quantity on the left hand side of the latter inequality is positive, taking into account that squaring both sides of that relation does not change the sense of the inequality, and that a multiplication by a negative factor (like $C$) instead reverses it, we can re-write the condition provided by the cosmological holographic principle as:
\beq
{\mathcal R} -4 \left( 1+ \frac{16 \pi^2 C}{\alpha^2} \right ) h(t) - \left( 1 + \frac{32 \pi^2 C}{\alpha^2}\right) \epsilon  > 0\,.
\eeq  
Thus, regardless of the location of the observer within such a universe, the bag energy of the cosmic fluid is constrained according to
\beq
C < \frac {\alpha^2 ( {\mathcal R} -\epsilon -4 h(t))}{ 32 \pi^2 (\epsilon + 2h(t))}\,.
\eeq
The condition that $C$ must be negative is automatically fulfilled, because it would require 
\beq
h(t) > -\frac{1 + 16AB}{4},
\eeq
which is automatically guaranteed for positive $A$ and $B$.

\subsection{Cosmological Constraints From the Second Law of Thermodynamics}\label{law}

In this subsection we will establish which relationships between the cosmological parameters characterizing the spacetime described by metric (\ref{eq1}) are compatible with the second law of thermodynamics. In agreement with standard physics, we will impose a time-increasing matter entropy for the cosmic fluid, and show that the further constraints among the free model parameters which would be derived do not contradict the ones already obtained. Therefore, in the class of models we are investigating, it is not necessary to weaken the second law of thermodynamics into the so-called {\it generalized second law} which requires only the sum of the matter entropy and of the gravitational entropy not to decrease during the cosmological evolution. This latter modification was needed for preserving the physical applicability of a number of cosmological models based on a Friedmann metric supported by radiation \cite{gsl1,gsl2,gsl3,gsl4,gsl5,gsl6}, a mixture of radiation and cosmological constant or a pressureless dark matter \cite{gsl7}, even beyond general relativity implementing torsion \cite{gsl8,gsl9} and braneworld \cite{gsl10} modifications. However, since we want to check the Weyl curvature hypothesis, it is most convenient to impose the second law on the matter sector, so that if the gravitational entropy (measured in some way by the square of the Weyl curvature) does indeed decrease -- which it does -- we can still have the possibility that the generalized second law may hold, from the matter contribution (otherwise we may rule out this cosmology as thermodynamically unphysical). 

Although we will be returning to the issue of gravitatrional entropy later on, it is worth emphasing already at this point that unlike in the case of stationary black holes \cite{ent1},  there is no agreement on a commonly accepted definition of gravitational entropy in cosmology. For example, the definition of cosmological entropy from the Weyl tensor as $S=C_{\alpha \beta}{}^{\gamma \delta}C_{\gamma \delta}{}^{\alpha\beta}$
fails when isotropic singularities occur \cite{ent2}. Moreover, a normalized gravitational entropy of the form  $S=C_{\alpha \beta}{}^{\gamma \delta}C_{\gamma \delta}{}^{\alpha\beta}/(R_{\alpha}{}^{\beta} R_{\beta}{}^{\alpha})$, while addressing the previous issue, clearly diverges in vacuum \cite{ent3}, just to mention the limitations of a couple of approaches in the literature. Instead, in this section we will show that our estimates on the size and age of the Universe are not affected by these uncertainties because just imposing a monotonically time-increasing matter entropy we can derive further realistic properties of the spacetimes under investigation.

From (\ref{cubic}) a time-increasing matter entropy would imply a time-increasing radius of the dynamical apparent horizon:
\beq
\frac{\d S_{\rm m}}{\d t}>0 \qquad  \Rightarrow \qquad \dot r_{\rm AH}>0 \,.
\eeq
Then, using (\ref{apparenthor}), the time evolution of the location of the dynamical apparent horizon can be computed explicitly as:
\beq
\dot r_{\rm AH}= - \frac{({\mathcal R} + \epsilon) \dot h(t)}{2C \, r_{\rm AH}\, (\epsilon + 2h(t))^2}\,,
\eeq
which gives the following inequality for accounting for the second law of thermodynamics\footnote{Remember that $C$ is negative, and that a multiplication by a negative factor switches the sense of the inequality.}:
\beq
({\mathcal R} + \epsilon) \dot h(t) >0\,.
\eeq
Implementing (\ref{hyperbola}) and using (\ref{HH}), we can conclude that the second law of thermodynamics \emph{requires an expanding universe} (i.e. with a positive Hubble function) in this model. A sharper condition would be:
\beq
A e^{ct} - B e^{-ct}>0\,,
\eeq
which imposes a lower limit on the size of the Universe
\beq
h(t)> 2B e^{-ct} \,,
\eeq
or equivalently on its age
\beq
\label{timeset}
t>\frac{1}{2c}\ln \frac{B}{A}\,.
\eeq
\emph{This is the range of the validity of the model imposed by the second law}, despite the model comes without a Big Bang singularity\footnote{One could of course entertain the possibility that the second law can somehow be violated, so that the model can be extrapolated back in time to the infinite past, with the entropy being decreasing up to a certain point. Such a scenario has been contemplated, e.g., in the context of bouncing cosmology \cite{Clutton-Brock,1104.1733}. The arrow of time problem would then require one to explain why the entropy shrinks down to such a small value during the bounce. See, however, \cite{1603.05834}. Alternatively we can impose the second law and take the more pragmatic viewpoint that for time earlier than the inequality (\ref{timeset}), the spacetime should be described by some other metric.}. 

Therefore, the strength of the Weyl curvature (\ref{weylcurv}) is decreasing with time because $\dot h(t)>0$ (its sign instead is arbitrary without any physical meaning  by definition \cite{NPC}). 
Our analysis suggests that in this class of models, a cosmological entropy defined as the square of the Weyl curvature would be decreasing during the evolution of the Universe, with the matter entropy being in charge of preserving the generalized second law of thermodynamics, as we will discuss more in detail in the next section. Interestingly, the information about the size and age of the Universe we have derived by imposing the second law of thermodynamics for the spacetime (\ref{eq1}) is not affected by the position of the observer, unlike in case of the Stephani universe \cite{stephani}.  The strength of the bag energy quantified by the parameter $C$ in the equation of state of the cosmic fluid (\ref{eos}) is not restricted by the the second law of thermodynamics either.

\section{Shearing Spacetime and the Violation of the Weyl Curvature Hypothesis}\label{sect4}

The \emph{Clifton-Ellis-Tavakol entropy} \cite{form7} (see also \cite{1405.0403} for more explanations) is a concrete realization of the general idea of Weyl curvature hypothesis,  which however does not depend only on the strength of the Weyl curvature, but also on the magnitude of the spacetime shear. It
appears to be a valuable proposal for a measure of the gravitational entropy because it increases monotonically during the formation of cosmic structures, that is when a gravitational collapse occurs \cite{form8,form9}. Moreover, the Clifton-Ellis-Tavakol proposal comes with many desirable features of a measure of entropy, because it is always non-negative, it vanishes in  -- and only in -- conformally flat spacetimes; it measures the strength of the local anisotropies of the gravitational field; and it can reproduce the Bekenstein-Hawking entropy of a black hole. The sturdiness of such proposal has been investigated explicitly in the inhomogeneous dust Lema\^itre-Tolman-Bondi universe and in the formation of local cosmic voids of about $50 -100$ Mpc size \cite{form10,form11,form12}. More generally, an appropriate notion for the gravitational entropy should be adopted for tracking the formation of cosmic structures because we know from statistical mechanics that the entropy is nothing else than an estimate of how many different microstates can realize the same macrostate, i.e. how many different inhomogeneous configurations on small scales are compatible with the dynamics of the same homogeneous universe after appropriate coarse graining \cite{form13}. 

In our spacetimes (\ref{eq1}), the so-called \lq\lq gravitational energy" \cite{form7}
\beq
\label{eng}
\rho_{\rm grav} = \frac{16 \pi G} {c^4} |\Psi_2|
\eeq
is \emph{decreasing} in time for an expanding universe with $H>0$; in particular this is the case compatible with the second law of thermodynamics, as previously discussed;  this behavior was examined in Fig. \ref{fig1}. We remark that the gravitational energy does not depend on the chameleon properties of the cosmic fluid, that is, on the parameter $C$.  Furthermore, according to the CET paradigm, the gravitational entropy of a Petrov D spacetime, like ours, depends not only on the gravitational energy, but also on the \lq\lq gravitational anisotropic pressure" which reads as \cite{form7,bolref}:
	\beq
	\label{anisg}
	\pi^{\rm grav}_{ab}=\frac{|\Psi_2|}{16 \pi G}(-x_a x_b +y_a y_b+z_a z_b +u^a u^b)\,.
	\eeq
	The spacelike unitary vectors which appear on the right hand side of this formula, and which constitute an orthonormal basis together with $u^a$, are given by:
	\beq
	x_a=\frac{1}{\sqrt{\epsilon +C r^2}}\partial_r \,, \qquad y_a=\frac{\sqrt{2}}{r\sqrt{\epsilon +2 h(t)}}\partial_\theta\,, \qquad z_a=\frac{\sqrt{2}}{r \sin \theta \sqrt{\epsilon +2 h(t)}}\partial_\phi\,.
	\eeq

Writing the Einstein equations in the so-called ``trace-reversed form"
\beq
R_{\mu\nu} = \frac{8 \pi G } {c^4} 	\left( T_{\mu\nu} - \frac12 g_{\mu\nu} T  \right)\,,
\eeq
we can easily compute
\begin{eqnarray}
R_{\mu\nu}R^{\mu\nu}&=&\left( \frac{8 \pi G }{c^4}\right)^2 \cdot T_{\mu\nu}T^{\mu\nu} \,=\, \left( \frac{8 \pi G }{c^4}\right)^2 \cdot [(\rho c^2)^2 + 3 p^2]   \\
&=& R^2 +18 C R + 108 C^2 \\
&=& 108 C^2 + \frac{4[\epsilon+{\mathcal R }+6Cr^2(\epsilon+2h(t))^2] \, [\epsilon+{\mathcal R }- 3 Cr^2(\epsilon+2h(t))^2]}{r^4 (\epsilon+2h(t))^4}\,.
\end{eqnarray}	
Thus, following the line of thinking of \cite{wcc3,buchert}, we can estimate the relative strength of the Ricci curvature with respect to the Weyl curvature (or equivalently of the ``matter energy'' vs. the gravitational energy):
\beq
\label{dominated}
\frac{R_{\mu\nu}R^{\mu\nu}} {\Psi_2^2} = \frac{36}{({\mathcal R} +\epsilon)^2} \Big[({\mathcal R} +\epsilon)^2 +3 ({\mathcal R} +\epsilon) C r^2 (\epsilon + 2 h(t))^2 +9 C^2 r^4 (\epsilon + 2 h(t))^4 \Big] \,,
\eeq
which would be constant and time-independent (equal to 36) for a non-chameleon cosmic fluid. The condition
\beq
\frac{R_{\mu\nu}R^{\mu\nu}} {\Psi_2^2} >1
\eeq
is equivalent to
\beq
35 ({\mathcal R} +\epsilon)^2 +108 ({\mathcal R} +\epsilon) C r^2 (\epsilon + 2 h(t))^2 +324 C^2 r^4 (\epsilon + 2 h(t))^4 \equiv [18 C r^2 (\epsilon + 2 h(t))^2  +3({\mathcal R}+\epsilon)]^2 +26({\mathcal R} +\epsilon)^2 >0
\eeq
which is trivially fulfilled. Thus, in the model of the universe (\ref{eq1}) the matter curvature dominates over the Weyl curvature all along the cosmic history. Unlike the cases of Bianchi and Lema\^itre-Tolman-Bondi universes with quantum initial conditions investigated in \cite{wcc3} our result is local and has not required the introduction of any ad-hoc averaging procedure.

\begin{figure}
		\begin{center}
			$
			\begin{array}{cc}
			{\includegraphics[scale=0.5, angle=0]{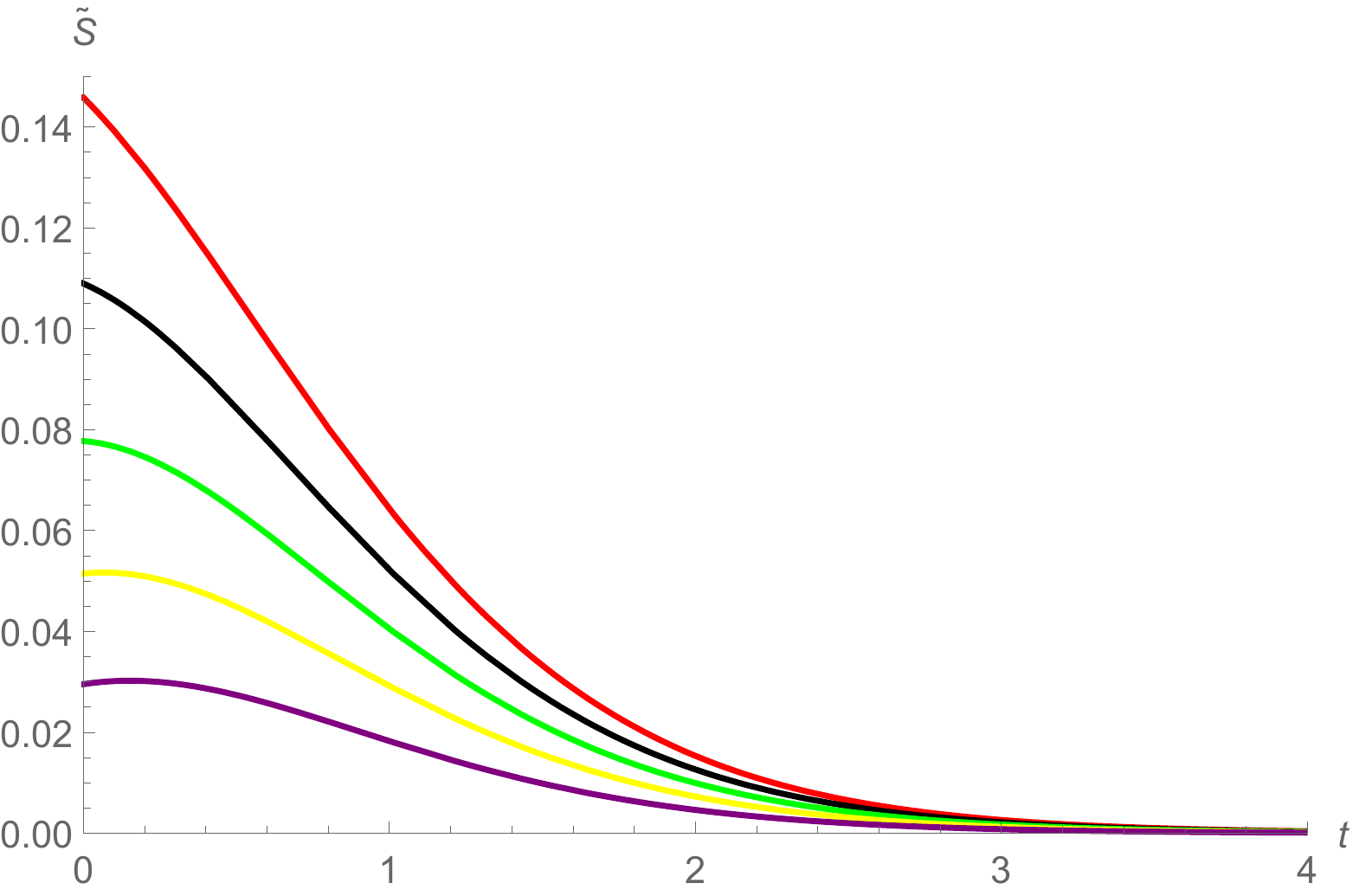}}
			\hspace{3mm}
			{\includegraphics[scale=0.5, angle=0]{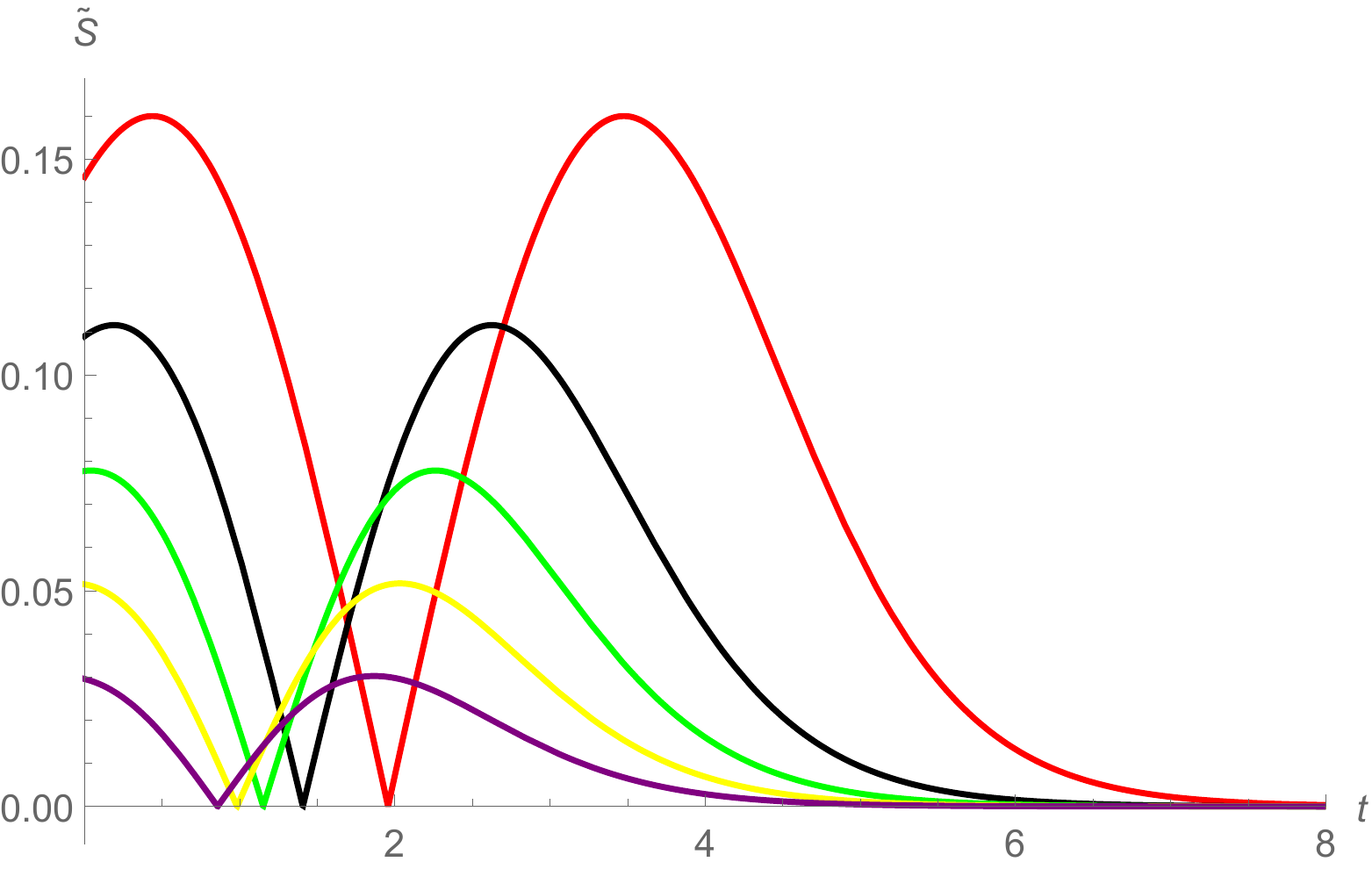}}\\
			(a) \hspace{90mm} (b) 
			\end{array}
			$
		\end{center}
		\caption{The figure depicts  the time evolution of the quantity ${\tilde S}:= T_{\rm grav} \dot s_{\rm grav}/dV$ computed from (\ref{tent}) for the universe (\ref{eqh3}) at the location $r=1$, and in units such that $c=1=8\pi G$. In panel (a) we choose A=0.50 and B=0.01 (red line),  B=0.03 (black line),  B=0.05 (green line),  B=0.07 (yellow line),  B=0.09 (purple line); in panel (b) we choose B=0.50 and A=0.01 (red line),  A=0.03 (black line),  A=0.05 (green line),  A=0.07 (yellow line),  A=0.09 (purple line). We remark that our choices for the numerical values of the free parameters are consistent with (\ref{hyperbola}). In Sec. \ref{law} we have showed that our model is physically acceptable in light of the second law of thermodynamics only at times $t>\frac{1}{2c}\ln\frac{B}{A}$, for which we can see that the gravitational entropy would be monotonically increasing (because its first time derivative is positive).
		}
		\label{fig3}
	\end{figure}

 The ``temperature'' of the free gravitational field is \cite{form7}
\beq
T_{\rm grav} = \frac{|u_{a;b}l^a n^b|} {\pi}=\frac{c^3 r} {8 \pi \sqrt{C r^2 + \epsilon}} \,.
\eeq
Interestingly, we note that for the choices $\epsilon=0$, and $\epsilon=-1$, the gravitational entropy is ill-defined because $C<0$, as from Eq.(\ref{rhoinf}), even in the case in which the chameleon properties of the cosmic fluid are suppressed for $C\to 0^-$. On the other hand, if we consider the ideal fluid limit together with $\epsilon=1$ we can eliminate this latter parameter by re-absorbing it into $r$. We stress that this is indeed in agreement with the discussion about the Lorentzian signature of the spacetime metric below  Eq.(\ref{pressure}).  The rate of evolution of the density of gravitational entropy according to the CET proposal is \cite{form7,bolref}
\beq
T_{\rm grav} \dot s_{\rm grav} = - \d V \sigma_{ab} \left(\pi^{ab}_{\rm grav} +\frac{(\rho c^2 +p)}{3 \rho_{\rm grav}} E^{ab}\right)     \,, 
\eeq
 where 
\beq
\d V= \frac{r^2 \sin\theta (\epsilon +2 h(t))}{2 \sqrt{C r^2 + \epsilon}}\, \d r \d\theta \d\phi 
\eeq
is the elementary volume form, in which we remind the reader that $\epsilon +2 h(t)>0$. 
Thus, the rate of evolution of the density of gravitational entropy for the universe (\ref{eqh3}) is explicitly given by:
\beq
\label{tent}
T_{\rm grav} \dot s_{\rm grav} = dV \frac{64 G \pi c^2 \dot h(t) (1-16AB)}{3(2h(t) +\epsilon)^3 r^3}\,,
\eeq	
where we have used Eqs. (\ref{eos})-(\ref{eng})-(\ref{anisg}). Thus, the gravitational entropy is increasing in time, because its first derivative is positive, thanks to the conditions (\ref{hyperbola}) and $\dot h(t)>0$. This result is in agreement with the {\it Gravitational Entropy Conjecture} according to which {\it Any universe that generates gravitational entropy cannot belong to a family of spatially homogeneous and isotropic FLRW models} \cite{bolref}. Furthermore, we remark that our result is exactand not approximated because it is not based on any perturbation theory, unlike the one in \cite{bolref}, of which it can be interpreted as an extension.  
 We display in Fig. \ref{fig3} the time evolution of the quantity ${\tilde S}:= T_{\rm grav} \dot s_{\rm grav}/dV$ for the universe (\ref{eqh3}) at the location $r=1$, and in units such that $c=1=8\pi G$. In panel (a) we choose A=0.50 and B=0.01 (red line),  B=0.03 (black line),  B=0.05 (green line),  B=0.07 (yellow line),  B=0.09 (purple line); in panel (b) we choose B=0.50 and A=0.01 (red line),  A=0.03 (black line),  A=0.05 (green line),  A=0.07 (yellow line),  A=0.09 (purple line). We remark that our choices for the numerical values of the free parameters are consistent with (\ref{hyperbola}). In Sec. \ref{law} we have showed that our model is physically acceptable in light of the second law of thermodynamics only at times $t>\frac{1}{2c}\ln\frac{B}{A}$, for which the gravitationally entropy would be monotonically increasing (because its first derivative is positive).
	
From Eq.(\ref{shear}), and regardless of the particular choice of the value of $\epsilon$,  we obtain the time-evolution of the shear as:
\beq
\frac{\d \sigma^2} {\d t} =  \frac{4 c^2 \, \dot h(t)  \, (2 h(t) - {\mathcal R} )} {3 r^2 (\epsilon + 2h(t))^3}  \,.
\eeq
Thus,  $\sigma^2$ is monotonically increasing for the case $\epsilon=1$,  should the universe be expanding (recalling that ${\mathcal R}<0$ for a well-defined Lorentzian signature with both $A$ and $B$ positive),  as it is its corresponding CET entropy, although the Weyl curvature and consequently the \lq\lq gravitational energy" are decreasing.  The fact that the shear is non-time-decreasing is an important difference than the current concordance model of cosmology. \cite{coleydem} has argued that a non-standard evolution of the shear may have indeed occurred at some stage of the evolution of the Universe because  the standard model of cosmology is in tension with the observed existence of certain primordial astrophysical structures. In fact, the sizes of the Large Quasar Groups are about 70-350 Mpc, despite the assumption of homogeneity on scales above 150 Mpc made within the standard model of cosmology. To be more specific, the catalogue  DR7QSO of the SDSS has identified at redshift $z\sim 1.27$ a specific Large Quasar Group characterized by a size of about 500 Mpc. Furthermore, it is conceivable that such structures can constitute the seeds for the formation of the cosmic filaments and walls \cite{lqg1,lqg2,lqg3,lqg4,lqg5}.

A few different varieties of cosmological models were recently studied \cite{1912.01414}, which shows that the  Clifton-Ellis-Tavakol gravitational entropy does start from zero at the Big Bang and monotonically grows afterwards in those models. Mathematically, the crucial difference in our case is due to the fact that the gravitational temperature is a constant in time, whereas in the examples studied in \cite{1912.01414}, both $\rho_{\rm grav}$ and $T_{\rm grav}$ diverge in the limit $t \to 0$, allowing the divergences to be canceled in a way such that the gravitational entropy vanishes. This does not happen in our case due to the gravitational temperature being time-independent. In fact, for $\epsilon=1$, the gravitational entropy is never singular in time. Thus we have demonstrated that in an accelerated expanding universe in which shear continues to grow, the CET gravitational entropy is increasing nevertheless the Weyl curvature is decreasing. This seems to be violating the Weyl curvature hypothesis because the definition of the gravitational entropy we have adopted depends not only on the Weyl curvature, but also on the shear tensor which we have showed to be in charge of make it increase .

\section{Discussion: What Is Gravitational Entropy?} \label{sect5}

Cosmological models describing the evolution of the primordial Universe may rely on three parameters, with the one based on the Shan-Chen fluid picture being one example \cite{sc}. Cosmological models accounting for the late-time dynamics of the Universe may rely on even more arbitrary quantities; for example the models with interaction in the dark sector may require also five free parameters (one for the equation of state of dark matter, two for accounting for a dynamical equation of state for dark energy, and two more which parametrize the energy exchanges between the two dark fluids) \cite{int1,int2,int3,int4,int5,int6}.

In this report we have assumed an inhomogeneous spherically symmetric spacetime admitting anisotropic shearing effects and whose evolution is driven by a stiffened fluid as a possible model for the early Universe. Our proposal is based on some mathematical solution of the Einstein's field equations found by Leibovitz-Lake-Van den Bergh-Wils-Collins-Lang-Maharaj. The three parameters of the model are constrained together with its topology by imposing a positive energy density for the cosmic fluid and analyzing the evolution of the matter entropy. In fact, according to the cosmological holographic principle the matter entropy inside a region bounded by the dynamical apparent horizon should be smaller than the area of that region (in the Planck units), while the more well-known second law requires a non-decreasing entropy for a universe with a physically realistic evolution. Therefore, we could evaluate the ``bag energy'' of those universes, their age, and their size. A negative deceleration parameter, and a time-decreasing Weyl curvature are obtained without the need of imposing any further condition. Despite being inhomogeneous, the location of the observer does not affect those estimates, unlike the case of the Stephani universe \cite{stephani}. Moreover, the effects of the Weyl curvature has been explored in light of the role of gravitational entropy in the formation of primordial cosmological structures; this analysis was not possible in the Stephani spacetime which is conformally flat.

Curiously, we have found the Weyl curvature -- {but not} the Clifton-Ellis-Tavakol gravitational entropy -- is monotonically decreasing as the universe expands, although the spacetime shear, and therefore the anisotropies, is increasing (corresponding to some sort of structure formation). This is despite the fact that we have constrained the model parameters with physical requirements that the matter (massless scalar) field must have positive energy density, must satisfy the second law of thermodynamics, and must satisfy the cosmological holographic principle. 
We emphasize that this behavior persists even if we switch off the chameleon property of the scalar field.
This behavior is surprising, since previous investigations have checked that {both the Weyl curvature and } the CET gravitational entropy are indeed increasing in time in a variety of cosmological models \cite{1912.01414}. What conclusions can be drawn from this? 

Unlike matter entropy, gravitational entropy is a tricky notion. This is partly due to the fact that we do not know what is the underlying ``atoms'' of gravitational degrees of freedom \cite{1012.4476}. 
A useful definition of gravitational entropy must, at the very least,
recover the Bekenstein-Hawking entropy of a black hole. It has been a longstanding problem as to what the Bekenstein-Hawking entropy is actually an entropy \emph{of} (and why would adding charge or rotation \emph{decrease} the entropy, compared to a neutral non-rotating black hole of the same mass?). This is not the only problem, however. Black holes have a lot more entropy than a typical matter configuration of the same size and energy. The former has $S \sim A$ (in Planck units), whereas the latter has only $S \sim A^{3/4}$ \cite{0908.1265v1, hara}. Consequently, as a star of mass $M$ collapses into a black hole, its entropy increases by a staggering factor of $10^{20}(M/M_{\odot})^{1/2}$, where $M_{\odot}$ denotes a solar mass \cite{hara}. Perhaps the process of collapse involves a huge increase in gravitational entropy, for whatever reason yet to be fully understood. 

If the notion of gravitational entropy is a good one\footnote{Wallace has argued that gravitational entropy is irrelevant in most contexts except in black hole physics, and that it suffices to consider the dynamics caused \emph{by} gravitational interactions \cite{0907.0659v1}. For our purpose, the mathematics is clear: Weyl curvature decreases in time. Whether this is really a measure of entropy caused \emph{by} gravity acting on matter, or entropy \emph{of} gravity, requires a deeper scrutiny. Essentially, this has to do with the decomposition of the Riemann curvature tensor into Ricci and Weyl part --  the latter remains, in part, free because its value is not provided by the field equations, but only some constraints must be accounted for through the Ricci identities. Thus one may say that the Weyl tensor constitutes the ``gravitational'' or ``geometrical'' part of the theory. See \cite{1902.05565} for further discussions and implications.}, then it should start small or even zero at the Big Bang, and then monotonically grow as structures like stars and galaxies and eventually black holes formed. In other words, it should, in some way, measure the increase in anisotropy\footnote{At least during the matter domination epoch. Thereafter, the Universe becomes dark energy dominated, and  eventually even black holes would Hawking evaporate away, though the \emph{total} entropy of the Universe, including that of all the Hawking quanta, should remain increasing.}. Since Weyl curvature increases during structure formation, it makes sense to define gravitational entropy such that said quantity does indeed increase. One of the most promosing candidate is the Clifton-Ellis-Tavakol gravitational entropy. Nevertheless, we have seen that inhomogeneities that arise together with a nontrivial spacetime shear, can correspond to a decrease in the Weyl curvature, {but not of} the CET gravitational entropy. This could indicate that we need to have a better definition for gravitational entropy, and the role of Weyl curvature in gravitational entropy should also be further revised. That is to say, the relation between gravitational entropy and spacetime shear might not be so straightforward.
Furthermore, in view of the Bekenstein-Hawking entropy of a black hole can be interpreted as entanglement entropy \cite{9401070, 9401072,0603081,0806.0402,1104.3712}, perhaps the role of entanglement entropy should also be considered \cite{1212.1087}. On the other hand, perhaps other notions of ``geometrical entropy'', such as the ``creation on a torus'' scenario that makes use of deep results in global differential geometry \cite{0611088v3, 0711.1656v2} is more useful to explain the initial low entropy state of the Universe. 
 
Finally, let us remark that, irrespective of the arrow of time issue, our paper fits inside the wider cosmological research which is trying to provide a critical assessment of the astrophysical datasets. In fact, many drawbacks of the $\Lambda$-Cold Dark Matter model, like the Hubble tension, the coincidence problem, and even the predicted existence of dark energy,  may be caused by interpreting the cosmological data which have been refined already implementing the Copernican principle \cite{crit1,crit2,crit3}. Our results on the other hand are based on theoretical considerations, which should complement observationally obtained constraints.

\subsection*{Acknowledgement}
Y.C.O. thanks NNSFC (grant No.11922508 \& No.11705162) and the Natural Science Foundation of Jiangsu Province (No.BK20170479) for
funding support. D.G. acknowledges support from China Postdoctoral Science Foundation (grant No.2019M661944). The authors thank Brett McInnes and Timothy Clifton for useful comments and suggestions.


\end{document}